\documentclass[twocolumn]{aa}
\usepackage{natbib}
\bibpunct{(}{)}{;}{a}{}{,} 
\usepackage{graphicx}
\usepackage{txfonts}

\def\eg{{e.g.\ }}
\begin{document}
\title{The VIMOS-VLT Deep Survey}
\subtitle{The evolution of galaxy clustering per spectral type to $z\simeq1.5$ 
  \thanks{based on data
    obtained with the European Southern Observatory Very Large
    Telescope, Paranal, Chile, program 070.A-9007(A), and on data
    obtained at the Canada-France-Hawaii Telescope, operated by
    the CNRS of France, CNRC in Canada and the University of Hawaii
    and observations obtained with MegaPrime/MegaCam, a joint project
    of CFHT and CEA/DAPNIA, at the Canada-France-Hawaii Telescope (CFHT)
    which is operated by the National Research Council (NRC) of Canada,
    the Institut National des Science de l'Univers of the Centre National
    de la Recherche Scientifique (CNRS) of France, and the University of
    Hawaii. This work is based in part on data products produced at TERAPIX
    and the Canadian Astronomy Data Centre as part of the Canada-France-Hawaii
    Telescope Legacy Survey, a collaborative project of NRC and CNRS.
  }
}
\author{
  B. Meneux      \inst{1}, 
  O. Le F\`evre  \inst{1}, 
  L. Guzzo       \inst{2}, 
  A. Pollo       \inst{1,2}, 
  A. Cappi       \inst{3},
  O. Ilbert      \inst{4},
  A. Iovino      \inst{5},
  C. Marinoni    \inst{5,6},
  H.J. McCracken \inst{7,8},
  D. Bottini     \inst{9},
  B. Garilli     \inst{9},
  V. Le Brun     \inst{1},
  D. Maccagni    \inst{9},
  J.P. Picat     \inst{10},
  R. Scaramella  \inst{11},
  M. Scodeggio   \inst{9},
  L. Tresse      \inst{1},
  G. Vettolani   \inst{12},
  A. Zanichelli  \inst{12},
  C. Adami       \inst{1},
  S. Arnouts     \inst{1},
  M. Arnaboldi   \inst{13},
  S. Bardelli    \inst{3},
  M. Bolzonella  \inst{4},
  S. Charlot     \inst{7,14},
  P. Ciliegi     \inst{3},
  T. Contini     \inst{10},
  S. Foucaud     \inst{15},
  P. Franzetti   \inst{9},
  I. Gavignaud   \inst{10,16},
  B. Marano      \inst{4},
  A. Mazure      \inst{1},
  R. Merighi     \inst{3},
  S. Paltani     \inst{17,18},
  R. Pell\`o     \inst{10},
  L. Pozzetti    \inst{3},
  M. Radovich    \inst{13},
  G. Zamorani    \inst{3}, 
  E. Zucca       \inst{3},
  M. Bondi       \inst{12},
  A. Bongiorno   \inst{4},
  G. Busarello   \inst{13},
  O. Cucciati    \inst{2,14},
  L. Gregorini   \inst{12},
  F. Lamareille  \inst{10},
  G. Mathez      \inst{10},
  Y. Mellier     \inst{7,8},
  P. Merluzzi    \inst{13},
  V. Ripepi      \inst{13},
  D. Rizzo       \inst{10}
}

\offprints{baptiste.meneux@oamp.fr}

\institute{
  Laboratoire d'Astrophysique de Marseille, UMR 6110 CNRS-Universit\'e de  Provence, BP8, 13376 Marseille Cedex 12, France
  \and
  INAF-Osservatorio Astronomico di Brera - Via Bianchi 46, I-23807, Merate, Italy
  \and
  INAF-Osservatorio Astronomico di Bologna - Via Ranzani,1, I-40127, Bologna, Italy
  \and
  Universit\`a di Bologna, Dipartimento di Astronomia - Via Ranzani, 1, I-40127, Bologna, Italy
  \and
  INAF-Osservatorio Astronomico di Brera - Via Brera 28, Milan, Italy
  \and
  Centre de Physique Theorique, UMR 6207 CNRS-Universite de Provence, case 907 F-13288 Marseille, France
  \and
  Institut d'Astrophysique de Paris, Universit\'e Pierre et Marie Curie, UMR 7095, 98 bis Bvd Arago, 75014, Paris, France
  \and
  Observatoire de Paris, LERMA, 61 Avenue de l'Observatoire, 75014 Paris, France
  \and
  INAF - IASF Milano, via Bassini 15, I-20133 Milano, Italy
  \and
  Laboratoire d'Astrophysique de l'Observatoire Midi-Pyr\'en\'ees (UMR 5572) - 14, avenue E. Belin, F31400 Toulouse, France
  \and
  INAF-Osservatorio Astronomico di Roma - Via di Frascati 33, I-00040, Monte Porzio Catone, Italy
  \and
  IRA-INAF - Via Gobetti,101, I-40129, Bologna, Italy
  \and
  INAF-Osservatorio Astronomico di Capodimonte - Via Moiariello 16, I-80131, Napoli, Italy
  \and
  Max Planck Institut fur Astrophysik, 85741, Garching, Germany
  \and
  School of Physics \& Astronomy, University of Nottingham, Nottingham NG7 2RD, England
  \and
  European Southern Observatory, Karl-Schwarzschild-Strasse 2, D-85748, Garching bei Munchen, Germany
  \and
  Integral Science Data Centre, ch. d'\'Ecogia 16, CH-1290 Versoix, Switzerland
  \and
  Geneva Observatory, ch. des Maillettes 51, CH-1290 Sauverny, Switzerland
  \and
  Universit\'a di Milano-Bicocca, Dipartimento di Fisica - Piazza delle Scienze, 3, I-20126 Milano, Italy
  \\              email: baptiste.meneux@oamp.fr
}

\date{Received ... 2005 ; accepted ... 2005}

\abstract{We measure 
  the evolution of  clustering for galaxies with different spectral types
  from 6495 galaxies with $17.5 \leq I_{AB} \leq 24$ and measured spectroscopic
  redshifts in the first epoch VIMOS-VLT Deep Survey (VVDS). We divide our
  sample into four classes, based on the fit of well-defined galaxy spectral
  energy distributions on observed multi-color data.
  We measure the projected correlation function $w_p(r_p)$ and estimate the best-fit
  parameters for a power-law real-space correlation function $\xi(r) = (r/r_o)^{-\gamma}$.
  We find the clustering of early-spectral-type galaxies to be markedly 
  stronger than   that of late-type galaxies at all redshifts up to $z\simeq1.2$.
  At $z\sim 0.8 $, early-type galaxies display a correlation length 
  $r_0=4.8 \pm$0.9~h$^{-1}$Mpc, while late types have $r_0=2.5 \pm $0.4~h$^{-1}$Mpc.
  For the latest class of star-forming blue galaxies, we are able
  to push our clustering measurement to an effective redshift $z\sim 1.4$,
  for luminous galaxies ($M_B(AB)\simeq -21$).  The clustering of
  these objects increases up to $r_0=3.42 \pm $0.7~h$^{-1}$Mpc for $z=[1.2,2.0]$. 
  The relative bias between early- and late-type galaxies within our magnitude-limited survey 
  remains approximately constant with $b=1.6 \pm 0.3$ from $z=0$ to $z=1.2$.
  This result is in agrement with the local findings and fairly robust against different
  way of classifying red and blue galaxies. When compared to the expected linear growth
  of mass fluctuations, a natural interpretation of these observations is that:
  (a) the assembly of massive early type galaxies is already mostly complete in the densest
  dark matter halos at $z\simeq1$;
  (b) luminous late-type galaxies are located in higher-density, more clustered regions of
  the Universe at $z\simeq1.5$ than their local low luminous counterpart, indicating that
  star formation activity is progressively increasing, going back in time, in the higher-density peaks that
  today are mostly dominated by old galaxies.


}

\authorrunning{B. Meneux et al.}

\titlerunning{VVDS: Clustering evolution per galaxy type}

\maketitle

%

\section{Introduction}

The measurement of the evolution of the clustering of galaxies is one the 
key elements to understand the evolution of the Universe and the
formation of the large-scale structures of the Universe. In the current paradigm, the
structures we observe today in the large-scale distribution of galaxies
are the result of the gravitational amplification of primordial
overdensities in the dark matter density field. Smaller scales  
are supposed to collapse first, with virialized structures as galaxies,
groups and clusters subsequently built up via multiple merging,
at a rate of growth essentially governed by the value of the mean
matter density $\Omega_M$. In this hierarchical formation scenario,
the evolution of the clustering of dark matter halos is now well
understood both analytically \citep{mowhite1996,sheth} and from N-body
simulations \citep{jenkins,kauffmann,springel}.

The visible component of galaxies is expected to assemble via the
collapse of baryonic matter within the dark matter halos. A naive
picture would let to conclude that galaxies follow the underlying
dark matter density field. 
The formation and evolution of galaxies is, however, driven by physical
processes such as cooling, star formation and feedback, which are
not easily linked
to the underlying dark matter halos \citep{kauffmann}.
Rather, a general prediction of most current models is that galaxy 
formation is favored in higher-density regions.  Objects forming
in regions of larger-than-average density collapse first and, 
looking back in time,  
are found to inhabit higher and higher density peaks (i.e., they
are more {\it biased}). Recent direct measurements of the evolution
of galaxy bias up to a redshift $z\sim1.5$ \citep{marinoni}  
provide supporting evidence to this biased picture of galaxy formation.

Although being well known since more that 25 years, the origin of
the relationship between galaxy morphological (or almost equivalently,
spectral) type and local structure is still not fully understood
\citep[\eg][]{dressler1980,smith2005,postman2005}.
Whether the higher fraction of early-type
galaxies in high- density regions is established {\it ab initio} (the
so-called {\it nature} hypothesis), or rather produced by
environmental effects during the life of the galaxy ({\it
nurture}) is still debated.   A practical effect of the
morphology-density relation is that 
early- and late-type galaxies trace differently the density field,
with the former class being more clustered (i.e. more {\it
biased}) than the latter.  In addition, more luminous galaxies (which
are preferentially early-types) are more clustered than less luminous
ones, with the two parameters -- luminosity and spectral type --
playing together in a subtle way
\citep{iovino93,madgwick,norberg,zehavi2002,loveday,guzzo1997,benoist1996}.  
Since the amount of biasing should be related to the mass and formation
history of a galaxy, one therefore expects the evolution of the clustering of
galaxies with different morphological types to contain important clues
on the assembly of the baryonic mass and the origin of galaxy types.

Large deep surveys consistently show that the clustering of 
the overall galaxy population evolves slowly with time
\citep{coil,olf_cf}.  Measurements from the VIMOS-VLT Deep Survey
\citep[VVDS,\ ][]{olf_cf}, indicate that the clustering amplitude of 
$L^\star$ galaxies increases by a factor $\sim 2.4$ from $z=1.5$ to
the present epoch. However, little is known about
the clustering evolution of different galaxy types
above $z\sim0.3$.  In the CFRS sample, no significant
difference was observed between the  clustering of red and blue
galaxies up to $z\sim1$ \citep{olfcfrs}, although the sample
was too small to reach firm conclusions.
Red galaxies at $z\simeq 0.6$ in the CNOC2 survey were
found to have a clustering amplitude $\sim5$ times higher than blue
galaxies \citep{carlberg97}.
Results from the CADIS survey using photometric redshifts indicate
that up to $z\sim1$ early spectral type galaxies are more clustered,
with a correlation amplitude  2.1 times larger than late type
at z=[0.75-1.07] \citep{phleps}.
A more recent and accurate estimate at $\left<z\right>=0.6$ \citep{phleps05}
confirms the same trend among red and blue galaxies with 
$r_0^{red}=5.39^{+0.30}_{-0.28}$~h$^{-1}$Mpc and
$r_0^{blue}=3.64^{+0.25}_{-0.24}$~h$^{-1}$Mpc.
At even higher redshifts, the measurement of type is substituted by
the selection of specific sub-populations, \eg \citet{daddi}
measure a strong
clustering of extremely red (J-K) galaxies at $2\leq z \leq 4$,
arguing that these may represent the progenitors of massive
elliptical galaxies at lower redshifts. 

In this paper we make an attempt to quantify accurately the contribution
of different galaxy types to the observed evolution of clustering
using the VVDS first epoch survey. The accuracy of
spectroscopic redshifts is used to project the correlation function
$\xi(r_p,\pi)$ in narrow velocity intervals and to measure the
correlation length $r_0$ and the slope $\gamma$ of the real space
correlation function as a function of redshift up to $z\sim 1.2$.

We describe the sample and our galaxy classification in Section 2,
and the method used to compute the real space correlation function
in Section 3. The clustering measurements are presented in Section 4,
and then discussed in Section 5.
We use a cosmology with $\Omega_M=0.3$, $\Omega_\Lambda=0.7$, 
throughout the paper. All distances, given in comoving units,
and absolute magnitudes are computed with $h=1$ (which denotes
$H_0/100~$km~s$^{-1}$~Mpc$^{-1}$).


\section{VVDS first epoch data}

\subsection{The sample}
We are using the VIMOS-VLT Deep Survey first epoch data
described in \citet{olf}. We concentrate on the 
VVDS-02h field, and we keep all galaxies with the
best redshifts available, with quality flags 2 to 9
(confidence level greater than 80\% in the redshift measurement).
AGN are excluded from the sample. The analyzed sample 
in this paper includes 6495 galaxies with spectroscopic
redshifts in an area of 1750~arcmin$^2$. 
The accuracy of the redshift measurement is
$\sim$~275 km.s$^{-1}$. A detailed description
of the sample is given in \citet{olf}.


\subsection{Galaxy classification}
\label{gal_class}

We used the galaxy classification introduced by \citet{zucca} 
to study the evolution of the luminosity function per galaxy type 
from the first epoch VVDS data. 
The classification has been obtained from the B to I band 
multi-wavelength dataset available in the VVDS-02h field \citep{olfim,
mccracken}, and the spectroscopic redshifts
\citep{olf}. The U band, which does not cover all the field \citep{radovich}
was not used in order to get a homogeneous classification on the sky.
Rest-frame colors have been 
matched to the empirical set of spectral energy distribution
(SED) templates described in \citet{arnouts99}, made of four 
observed spectra \citep[][CWW hereafter]{cww} and two starburst SEDs computed
with GISSEL \citep{bruzualcharlot93}.

\begin{figure}
  \includegraphics[width=9cm]{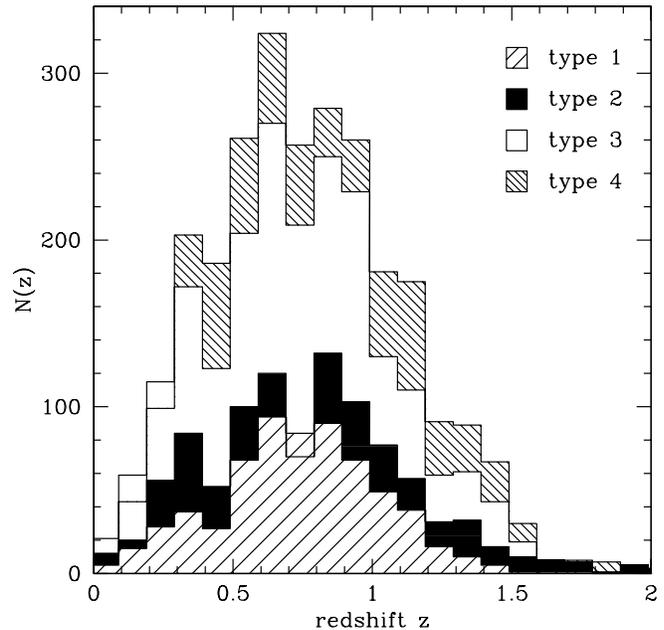}
      \caption{Redshift distribution of the four galaxy types }
      \label{nofz}
\end{figure}

\begin{figure}
  \includegraphics[width=9cm]{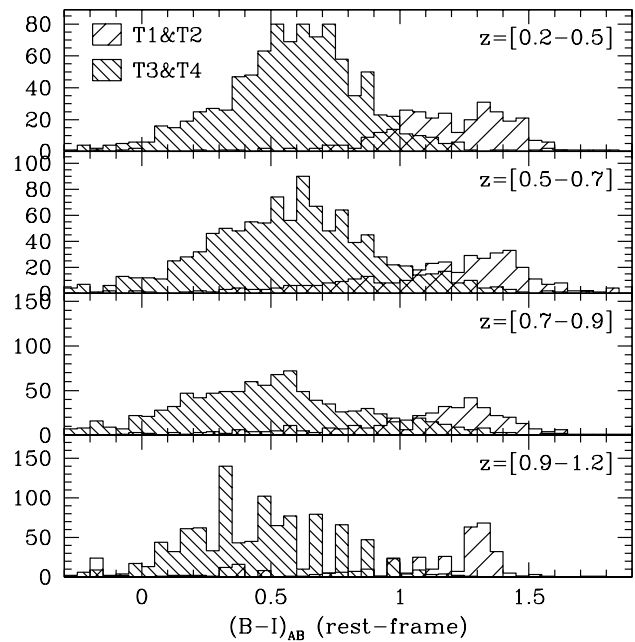}
      \caption{Rest-frame color $(B-I)_{AB}$ for early-type galaxies (type 1\&2) and late-type galaxies (type 3\&4)
        computed from the best fit of our multi-wavelength dataset with PEGASE.2 templates \citep{pegase2}.}
      \label{colorbi}
\end{figure}

The original spectroscopic sample of 6495 galaxies was divided into
four spectral classes from early-type (type 1) to 
irregular / star-forming galaxies (type 4) based on the CWW templates. 
The latest type also includes galaxies with SEDs well represented
by starburst templates. The properties of these four classes 
of galaxies are summarized in Table \ref{table_type} and
their redshift distributions are shown in Figure \ref{nofz}.
As a result of the VVDS I-band selection, we are increasingly loosing red early-type galaxies for
redshifts $z>\sim1.2$ (Figure \ref{colorbi}) because galaxies are
selected from their rest frame UV flux at these redshifts, but
late type galaxies continue to be identified up to the highest redshifts 
discussed in this paper.

\begin{figure}
  \includegraphics[width=9cm]{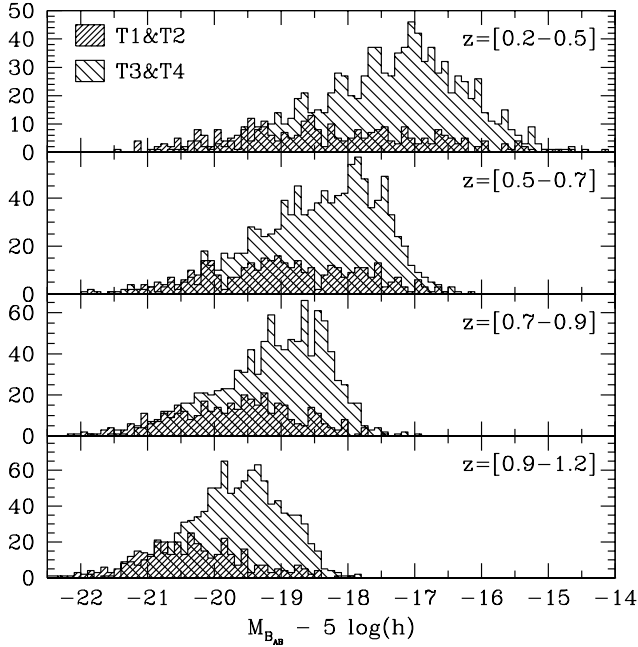}
      \caption{B-band luminosity distribution $M_B(AB)$ for early-type galaxies (types 1\&2) and late-type galaxies (types 3\&4)}
      \label{brest}
\end{figure}

As described in \citet{zucca},
the evolution of the global luminosity function in the VVDS
is strongly driven by type 4 galaxies, which evolve
by $\simeq$ -2 magnitudes from $z\sim1.5$ and present a steep
faint end luminosity function slope. Conversely, the luminosity
function of early type galaxies remains stable over this
redshift range. The relative distribution
of absolute magnitudes per type presented in Figure \ref{brest}
is the result of this evolution. 

\begin{table}
\caption{The four VVDS galaxy types}
\label{table_type}
\centering
\begin{tabular}{c c c c}
\hline\hline
type &                & number of &             indicated           \\
     &                & objects   &         rest-frame color       \\
\hline
   1 & E/S0           &     645   &  $1.3  < (B_{AB}-I_{AB})       $\\
   2 & early spiral   &    1004   &  $0.95 < (B_{AB}-I_{AB}) < 1.3 $\\
   3 & late spiral    &    2104   &  $0.68 < (B_{AB}-I_{AB}) < 0.95$\\
   4 & irregular      &    2742   &  $       (B_{AB}-I_{AB}) < 0.68$\\
     & starburst      &           &                                 \\
\hline
\end{tabular}
\end{table}

In summary, the type classification used in this paper is
a spectral classification based on a set of spectral 
templates covering from early type galaxies to late and star-forming
galaxies. Although it is tempting to associate spectral type to
morphological type, and a close relationship exists between them,
the spectral classification used in this paper is more closely related to a 
color classification from ``red'' to ``blue'' galaxies.


\section{Measurement and associated errors: method}

The method to recover the correlation length $r_0$ and the slope $\gamma$ 
of the real-space correlation
function is well detailed in the paper of \citet{pollo}. 
We are focusing here on the most critical 
aspects of the method when applied
to compute the clustering properties
as a function of galaxy type.

\subsection{The real space correlation functions parameters}

We computed the bi-dimensional two-point correlation function $\xi(r_p,\pi)$ 
for each galaxy type, in different redshift bins chosen to maximize 
the number of objects hence the signal to noise ratio of the
clustering measurement. This was estimated using the \citet{lansal} estimator

\begin{equation}
\xi(r_p,\pi) = \frac{N_R(N_R-1)}{N_G(N_G-1)} \frac{GG(r_p,\pi)}{RR(r_p,\pi)}
 - \frac{N_R-1}{N_G} \frac{GR(r_p,\pi)}{RR(r_p,\pi)} + 1
\label{lseq}
\end{equation}
where $N_G$ is the mean galaxy density (or, equivalently, the total
number of objects) in the survey; $N_R$ is the mean density of a
catalog of random points distributed within the same survey volume;
$GG(r_p,\pi)$ is the number of independent galaxy-galaxy pairs with
separation between $r_p$ and $r_p+dr_p$ perpendicular to the line-of-sight 
and between $\pi$ and $\pi+d\pi$ along the line of sight; 
$RR(r_p,\pi)$ is the number of independent
random-random pairs within the same interval of separations and $GR(r_p,\pi)$
represents the number of galaxy-random cross pairs.

The real-space correlation function $\xi(r)$ represents the excess 
probability for a given pair of galaxies to be observed at a separation 
$r$ \citep{peebles80}. It can be derived from $\xi(r_p,\pi)$
using the formalism of \citet{dp83}, computing $w_p(r_p)$, the projection 
of $\xi(r_p,\pi)$ along the line of sight.
\begin{equation}
w_p(r_p) \equiv 2 \int_0^\infty \xi(r_p,\pi) dy = 2 \int_0^\infty \xi\left[(r_p^2+y^2)^{1/2}\right] dy
\label{wpdef}
\end{equation}
In practice, the upper integration limit must be chosen finite,
as to include the real signal, without adding extra noise which is 
dominant above a certain $\pi$.  Based on direct tests with the
GalICS mock samples \citep{blaizot}, 
we choose a value $\pi_{max}=20$~h$^{-1}$Mpc. 
This is high enough to sum the real signal, without adding
too much noise. It is compatible with the VVDS velocity
measurement error of $\sim$~275 km.s$^{-1}$.
Assuming a power-law form for the real-space correlation function $\xi(r)=(r/r_o)^\gamma$, 
the integral in equation \ref{wpdef} can be solved analytically
in terms of Gamma functions, yielding a $w_p(r_p)$ which
is itself a power law.
The fitting method of the measured $w_p(r_p)$, detailed in \citet{pollo}, 
provides the values of the correlation length $r_0$ and the slope $\gamma$ 
with their associated errors.  
A specific treatment have been applied to 
compute the clustering properties per galaxy types, as described in Sections \ref{part_weight} and  
\ref{part_error}.

\subsection{Biases and weights for each spectral class}
\label{part_weight}

In order to correct for the observational biases introduced
by the VVDS observing strategy in the VVDS-02h field,
we used the same correction scheme explained in \citet{pollo}.
The main biases are the incompleteness in
redshift measurement and the complex geometry of the field. 
This results in a spatial sampling that changes as a function of
position in the field. To account for galaxies with unknown redshift 
(either unobserved or discarded because of a low redshift quality flag), 
the observed galaxies are given a weight.
The goal of this weight is to recover the true number of galaxy pairs at 
a given angular scale.

The computation of this local weight takes into account the local number 
of galaxies (with and without a redshift) around a given galaxy in the 
spectroscopic sample. The parent photometric sample used for
this computation must have the same properties as the spectroscopic one. 
For instance, the weight computed for the early type galaxies from the spectroscopic
sample must make a reference only to the early type galaxies in the
underlying complete photometric sample, without including
other galaxy types. In this approach, the parent photometric catalogue used 
to compute the weight must therefore only contain galaxies with the same
properties as those of the spectroscopic sample.

Photometric redshifts of all the galaxies in the VVDS-02h have 
been derived from BVRI photometry \citep{olfim,mccracken},
and ugriz bands from the CFHT Legacy Survey.
Details are presented in \citet{ilbert2}. The photometric redshifts and 
rest-frame magnitudes were then used to give a spectral type to all 
these galaxies, according to the method 
explained in Section \ref{gal_class}. The complete parent photometric 
catalogue was then divided in four catalogues, one for each type. 
Note that the photometric redshifts are only used to define a spectral
type to each galaxy in the whole photometric catalog in order to compute 
the weights associated to each galaxy type, in addition
to the spectroscopically measured galaxies. Photometric redshifts 
are not directly used here in the computation of the correlation 
function.

\subsection{Error measurement}
\label{part_error}

Even if our spectroscopic sample is the largest to date at this depth,
and despite the angular size of the VVDS-02h (1750~arcmin$^2$),
the uncertainties associated to the measurement 
of the power-law parameters of the real-space correlation function 
$\xi(r)$ are largely dominated by the cosmic variance. 

To assess this, we have constructed 50 VVDS-02h mock surveys from
the GalICS simulation \citep{blaizot}. GalICS is a semi-analytic
model for  galaxy evolution post-processed to a large cosmological
N-body simulation. These mock surveys are quasi-independent and are
built applying all the VVDS-02h observing biases: the complex geometry
of the field, the spectroscopic selection function, the redshift
measurement accuracy, the incompleteness
(see \citet{pollo} for more information).
We use the clustering variance in the mock
samples to quantify the cosmic variance expected on the correlation
length $r_0$ and the slope $\gamma$ of the real-space correlation
function.

Our method to assign a spectral type to a given galaxy does not work 
so well with the GalICS simulations. We were not able to recover a 
realistic redshift distribution for the four spectral types using
the CFH12k BVRI apparent magnitudes produced by GalICS. For example,
there were no Type 1 (elliptical) galaxies beyond z$\sim$0.5 in most of the
50 GalICS cones. Conversely, the galaxy type classification given
in for GalICS data do not match ours. This ``Morphological'' type
classification is based on rest-frame B-band (bulge+burst)-to-disc
luminosity ratio. This provides a huge ``spiral'' class (on average
$\sim$90\% of the full dataset), and two equivalent (in number of
objects) ``elliptical'' and ``lenticular'' classes. These two last
classes could match our type 1 class. But then, this is quite
difficult to divide the huge ``spiral'' class in our 3 later types.

Computing error measurements from a bootstrap resampling method
slightly under-estimates the real variance: if we consider the full
VVDS-02h data set \citep{olf_cf}, the relative errors computed
from GalICS are $\sim1.5-2$ times greater than error computed
from bootstrap resampling. We expect similar factor when dividing
our sample per spectral type.

Therefore the approach that has been chosen to compute measurement errors
consists in taking randomly,
in each of the 50 mock VVDS-02h surveys, a proportion of galaxies
equal to that found in the VVDS-02h spectroscopic sample for a given
spectral class. Even if the clustering measured is the one of the
``full'' population with higher incertainties, the statistic of each
``type'' is representative of the real one. The variance from
field to field is artificially increased by a broader range of
clustering properties in the simulated data, and therefore, this method 
is very conservative in the way that it is expected to produce
errors for the observed data larger than in reality.


\section{Clustering evolution for each galaxy type}

\subsection{Measurement of $r_0$ and $\gamma$}
\begin{figure}
  \includegraphics[width=9cm]{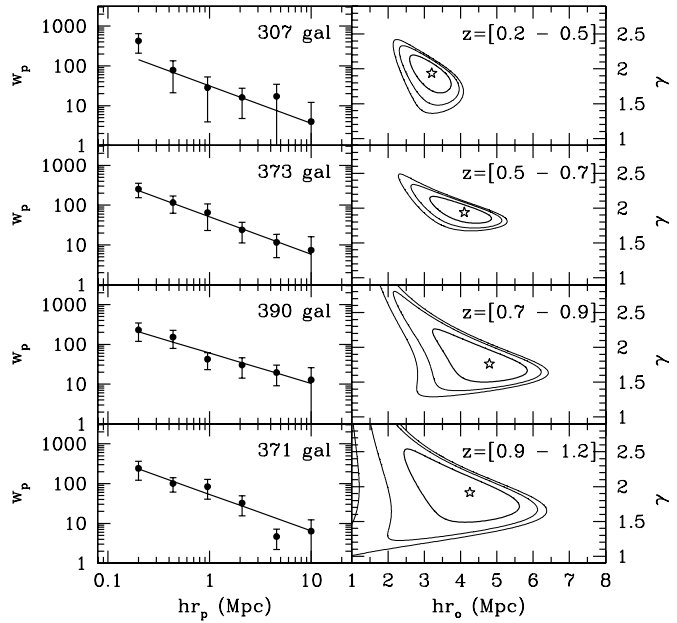}
      \caption{{\it (left)} Correlation function $w_p(r_p)$  as a function 
        of redshift  for early-type galaxies with types 1 and 2.
        {\it (right)} Error contours (68\%, 90\% and 95\% likelihood levels)
	associated to the measurement
        of the correlation length $r_0$ and slope $\gamma$
        of the correlation function.}
      \label{wp1}
\end{figure}

\begin{figure}
  \includegraphics[width=9cm]{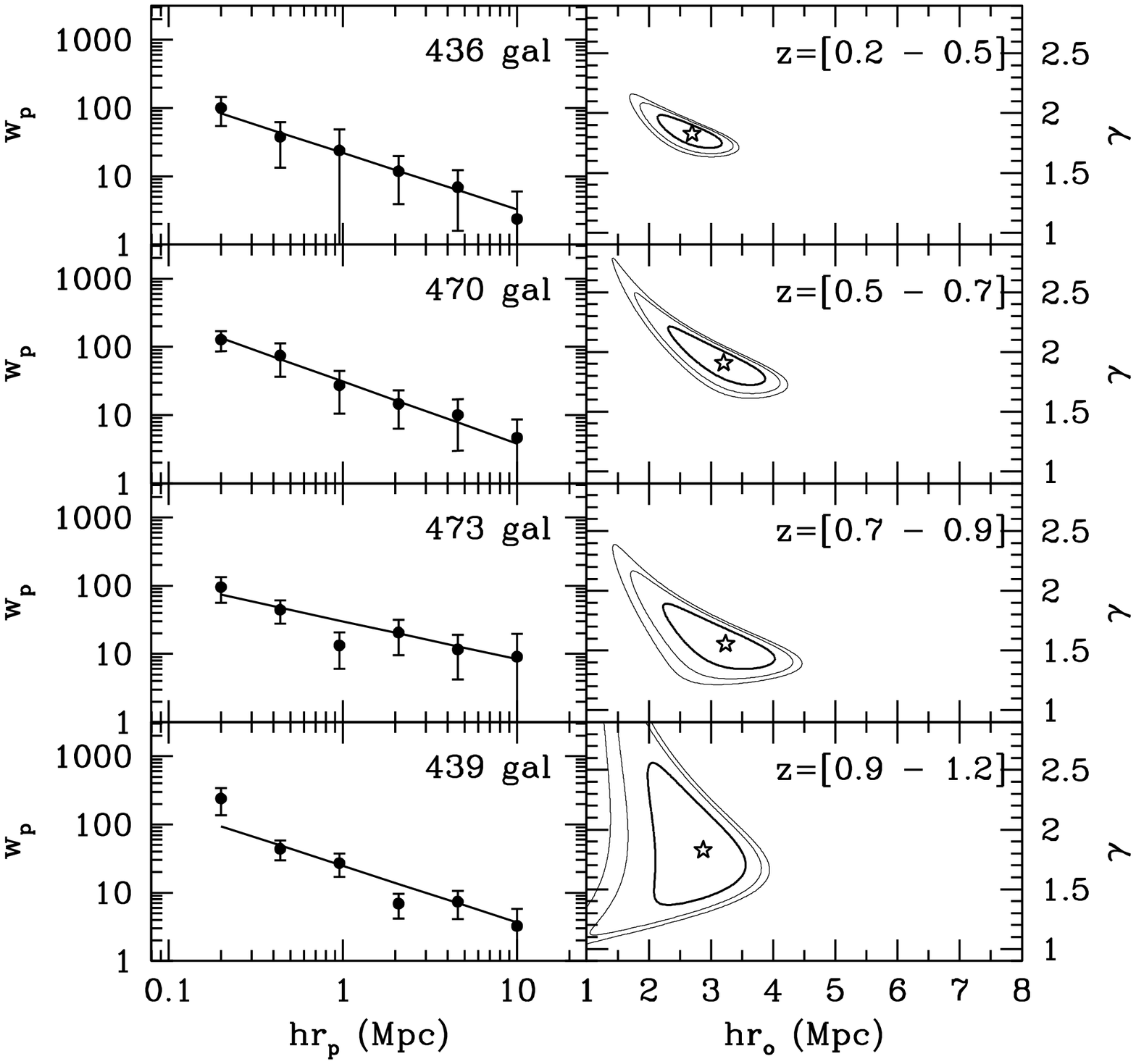}
      \caption{Same as Figure \ref{wp1} for galaxies with type 3.}
      \label{wp2}
\end{figure}

\begin{figure}
  \includegraphics[width=9cm]{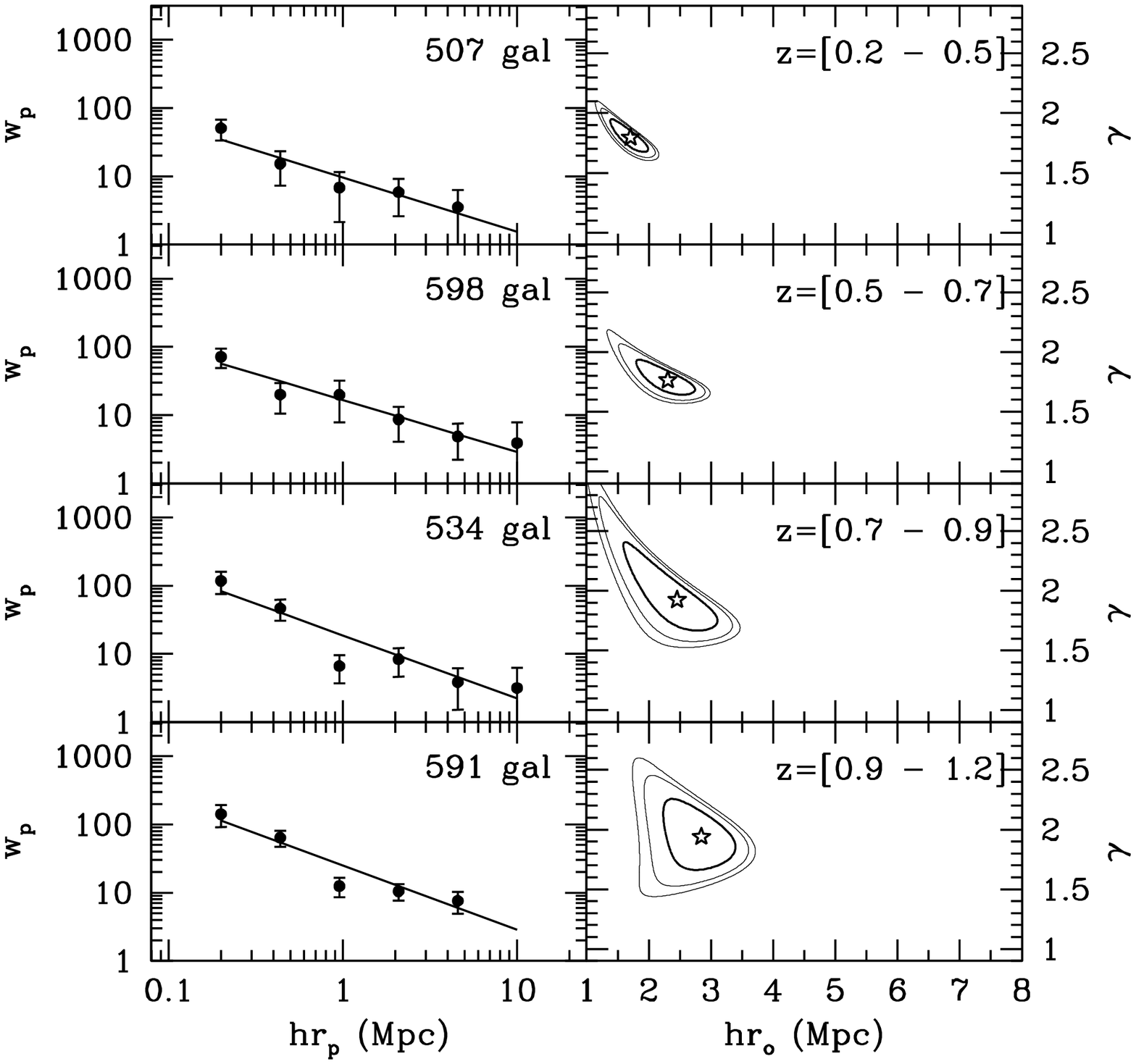}
      \caption{Same as Figure \ref{wp1} for galaxies with type 4.}
      \label{wp3}
\end{figure}

\begin{figure}
  \includegraphics[width=9cm]{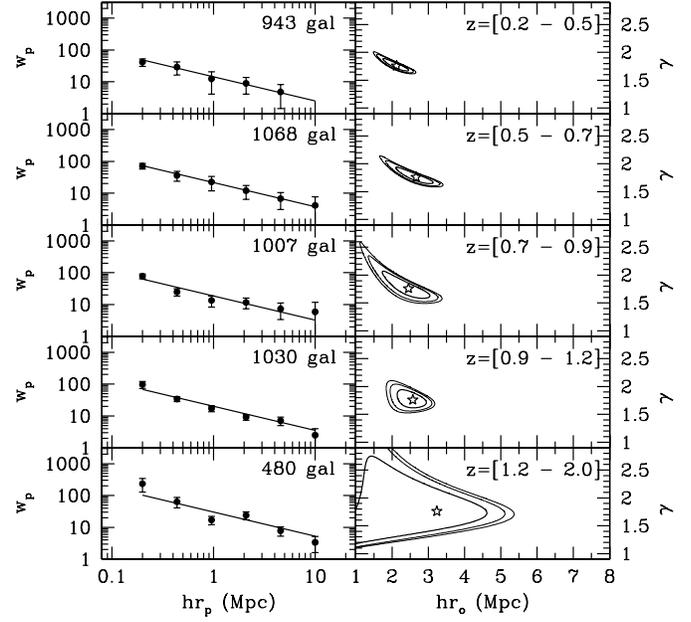}
      \caption{Same as Figure \ref{wp1} for galaxies with type 3 and 4.}
      \label{wp4}
\end{figure}

We have computed the correlation function $\xi(r_p,\pi)$ and its 
projection $w_p(r_p)$ in the VVDS-02h field, for each galaxy type of 
our classification, in increasing redshift slices. In order to improve
the clustering signal, and depending on the redshift bin
considered, we merged the two earliest-type galaxy samples 
$T1$ and $T2$ and the two latest types $T3$ and $T4$. Merging types
1 and 2 would clearly dilute the clustering signal of type 1
galaxies if they are more strongly clustered than type 2.
The measurements 
of $w_p(r_p)$ are presented in Figures \ref{wp1} to \ref{wp4}.

The projection $w_p(r_p)$ has been fitted by a power-law on scales 
$0.1 \le r_p \le 10$~h$^{-1}$Mpc. The values of the correlation 
length $r_0$ and the slope $\gamma$ of the real-space correlation function 
are summarized in Table \ref{table1_values}.
A clear difference in clustering strength is observed 
between the 3 classes of galaxies: T1\&2, T3 and T4.
We find that on average the
clustering strength of early type galaxies is 1.8 times larger than
the clustering strength of type 4 galaxies. This is observed
over the full redshift range $0.2 \leq z \leq 1.2$ and
this result is significant.
The probability that the correlation lengths of these two populations
are the same is indeed only 2.3\% for the redshift bin z=[0.2,0.5], 
2.7\% for z=[0.5,0.7], 2.2\% for z=[0.7,0.9], and 14.5\% for z=[0.9,1.2].

\begin{figure*}
\begin{center}
  \includegraphics[width=13cm]{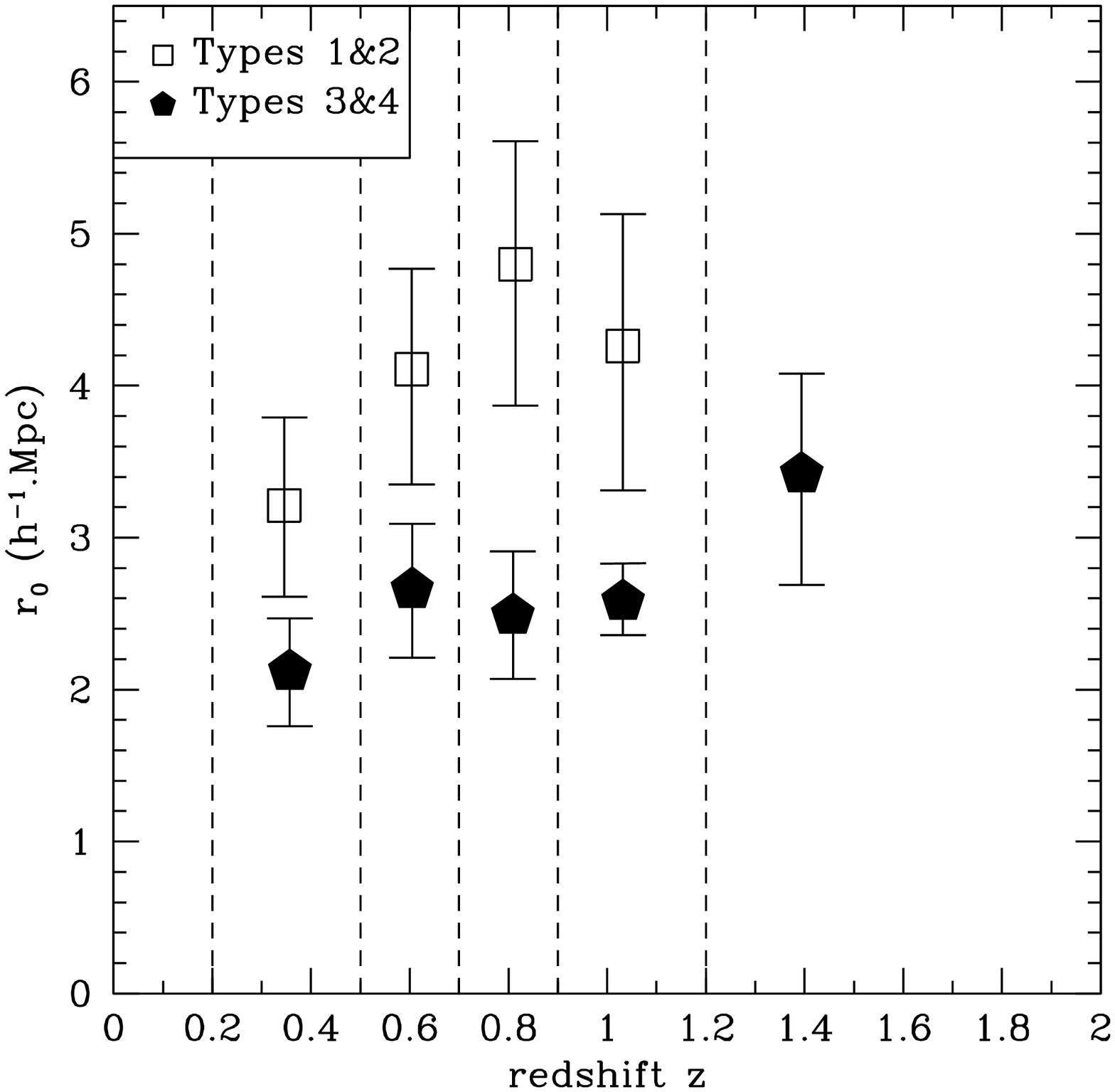}
  \caption{Correlation length $r_0$ as a function of redshift for early (T1\&2) and late (T3\&4) type galaxies.
    Here the slope $\gamma$ is a free parameter.}
  \label{evol_r0z}
\end{center}
\end{figure*}

The evolution of the clustering length for the different classes of galaxies 
is presented in Figure \ref{evol_r0z}.  An overall
increase in the clustering of all galaxy types is observed up to $z\simeq1$.
In the range z=[0.9,1.2], there is a slight indication for a
weakening in the clustering of early type galaxies.
The clustering length of late type / star-forming galaxies continues to rise up to
redshift $\sim 2$.

\begin{table*}  
  \caption{Measurement of the clustering length for galaxy subsamples 
    ranging from z=0.2 to z=2, letting the slope $\gamma$ free or
    fixing the slope to the average value measured for a given type.
    The absolute $M_{B_{AB}}$ magnitude and number of objects
    are indicated.} 
  \label{table1_values}   
\centering      
\begin{tabular}{l c c c c c c c l}      
\hline\hline    
   type & number of & redshift  & effective & \multicolumn{2}{c}{$M_{B_{AB}} - 5 log(h)$} &         $r_0(z)$        &        $\gamma$        &      $r_0(z)$ at fixed $\gamma$ \\ 
        & galaxies  & range     & redshift  &        mean     &      median    &      (h$^{-1}$Mpc)     &                        &             (h$^{-1}$Mpc)              \\ 
\hline          
     1  &     164   & [0.2-0.6] &   0.427   &      -18.799    &     -19.020    &  $3.14^{+0.76}_{-0.94}$ & $2.41^{+0.33}_{-0.37}$ & $3.51^{+0.65}_{-0.71}$  ($\gamma=2.15$) \\ 
        &     332   & [0.6-1.0] &   0.779   &      -19.755    &     -19.819    &  $4.35^{+0.68}_{-0.98}$ & $2.01^{+0.46}_{-0.29}$ & $3.93^{+0.55}_{-0.74}$                  \\ 
        &     237   & [0.8-1.2] &   0.964   &      -20.324    &     -20.347    &  $3.89^{+1.04}_{-1.44}$ & $2.04^{+0.80}_{-0.45}$ & $3.62^{+0.89}_{-1.26}$                  \\ 
\hline          
     2  &     299   & [0.2-0.6] &   0.409   &      -18.352    &     -18.419    &  $2.79^{+0.42}_{-0.44}$ & $1.88^{+0.17}_{-0.14}$ & $2.76^{+0.35}_{-0.32}$  ($\gamma=1.90$) \\ 
        &     420   & [0.6-1.0] &   0.796   &      -19.525    &     -19.503    &  $4.92^{+0.72}_{-0.77}$ & $1.82^{+0.27}_{-0.20}$ & $4.89^{+0.58}_{-0.58}$                  \\ 
        &     358   & [0.8-1.2] &   0.972   &      -20.059    &     -20.115    &  $4.35^{+0.96}_{-1.14}$ & $2.00^{+0.38}_{-0.26}$ & $4.62^{+0.85}_{-0.96}$                  \\ 
\hline          
     3  &     437   & [0.2-0.5] &   0.346   &      -17.478    &     -17.332    &  $2.69^{+0.50}_{-0.54}$ & $1.83^{+0.16}_{-0.11}$ & $2.73^{+0.44}_{-0.42}$  ($\gamma=1.78$) \\ 
        &     470   & [0.5-0.7] &   0.606   &      -18.665    &     -18.601    &  $3.21^{+0.60}_{-0.69}$ & $1.91^{+0.24}_{-0.15}$ & $3.33^{+0.54}_{-0.59}$                  \\ 
        &     473   & [0.7-0.9] &   0.810   &      -19.296    &     -19.212    &  $3.24^{+0.62}_{-0.65}$ & $1.56^{+0.22}_{-0.15}$ & $3.21^{+0.60}_{-0.60}$                  \\ 
        &     439   & [0.9-1.2] &   1.027   &      -19.843    &     -19.835    &  $2.88^{+0.44}_{-0.44}$ & $1.83^{+0.36}_{-0.27}$ & $2.82^{+0.49}_{-0.50}$                  \\ 
        &     191   & [1.2-2.0] &   1.394   &      -20.756    &     -20.751    &  $3.60^{+0.83}_{-1.38}$ & $1.73^{+0.35}_{-0.46}$ & $3.65^{+1.02}_{-3.04}$                  \\ 
\hline          
     4  &     507   & [0.2-0.5] &   0.366   &      -17.255    &     -17.170    &  $1.71^{+0.30}_{-0.31}$ & $1.80^{+0.16}_{-0.12}$ & $1.71^{+0.30}_{-0.30}$  ($\gamma=1.86$) \\ 
        &     598   & [0.5-0.7] &   0.604   &      -18.277    &     -18.136    &  $2.31^{+0.40}_{-0.40}$ & $1.76^{+0.14}_{-0.10}$ & $2.28^{+0.42}_{-0.39}$                  \\ 
        &     534   & [0.7-0.9] &   0.807   &      -18.944    &     -18.829    &  $2.46^{+0.49}_{-0.57}$ & $1.92^{+0.31}_{-0.19}$ & $2.49^{+0.53}_{-0.58}$                  \\ 
        &     591   & [0.9-1.2] &   1.036   &      -19.553    &     -19.494    &  $2.85^{+0.36}_{-0.31}$ & $1.94^{+0.18}_{-0.15}$ & $2.91^{+0.40}_{-0.37}$                  \\ 
        &     289   & [1.2-2.0] &   1.394   &      -20.441    &     -20.397    &  $3.30^{+0.64}_{-0.87}$ & $1.92^{+0.36}_{-0.32}$ & $3.27^{+1.01}_{-2.35}$                  \\ 
\hline          
  1\&2  &     307   & [0.2-0.5] &   0.346   &      -18.327    &     -18.512    &  $3.21^{+0.58}_{-0.60}$ & $1.94^{+0.27}_{-0.29}$ & $3.21^{+0.40}_{-0.40}$  ($\gamma=1.89$) \\ 
        &     373   & [0.5-0.7] &   0.604   &      -18.996    &     -19.020    &  $4.11^{+0.66}_{-0.76}$ & $1.94^{+0.19}_{-0.12}$ & $4.14^{+0.53}_{-0.57}$                  \\ 
        &     390   & [0.7-0.9] &   0.814   &      -19.719    &     -19.635    &  $4.80^{+0.81}_{-0.93}$ & $1.76^{+0.30}_{-0.18}$ & $4.41^{+0.62}_{-0.69}$                  \\ 
        &     371   & [0.9-1.2] &   1.032   &      -20.333    &     -20.383    &  $4.26^{+0.87}_{-0.95}$ & $1.92^{+0.31}_{-0.24}$ & $4.32^{+0.79}_{-0.90}$                  \\ 
\hline          
  3\&4  &     943   & [0.2-0.5] &   0.357   &      -17.358    &     -17.230    &  $2.12^{+0.35}_{-0.36}$ & $1.78^{+0.14}_{-0.09}$ & $2.12^{+0.32}_{-0.32}$  ($\gamma=1.76$) \\ 
        &    1068   & [0.5-0.7] &   0.605   &      -18.448    &     -18.321    &  $2.66^{+0.43}_{-0.45}$ & $1.74^{+0.15}_{-0.09}$ & $2.66^{+0.41}_{-0.41}$                  \\ 
        &    1007   & [0.7-0.9] &   0.809   &      -19.109    &     -19.008    &  $2.49^{+0.42}_{-0.42}$ & $1.67^{+0.17}_{-0.11}$ & $2.46^{+0.48}_{-0.50}$                  \\ 
        &    1030   & [0.9-1.2] &   1.032   &      -19.677    &     -19.626    &  $2.58^{+0.25}_{-0.22}$ & $1.86^{+0.11}_{-0.08}$ & $2.58^{+0.28}_{-0.25}$                  \\ 
        &     480   & [1.2-2.0] &   1.394   &      -20.566    &     -20.510    &  $3.42^{+0.66}_{-0.73}$ & $1.99^{+0.24}_{-0.20}$ & $3.24^{+0.90}_{-1.23}$                  \\ 
\hline          
\end{tabular}   
\end{table*}

\subsection{Relative bias and its evolution}

Since it is not possible to measure directly the real-space correlation 
function of the mass, any attempt to measure the bias $b$ from galaxy
samples has either to assume the knowledge of the background cosmological
parameters \citep[\eg][]{marinoni}, or to obtain the mass variance
from other observables like the Cosmic Microwave Background \citep[\eg][]{lahav}.

The simplest definition of 
the bias $b$ that links linearly the mass density field $\rho$ of the mass to
the galaxy density field $n$ is (for a fixed scale $R$ over which
$\rho$ and $n$ are measured)
\begin{equation}
\frac{\delta n}{\left<n\right>} = b \frac{\delta\rho}{\left<\rho\right>} \,\,\,\,   ,
\label{bias1}
\end{equation}
where $b$ is simply a number, independent on scale and on the local
value of $\delta\rho/\left< \rho \right>$
hypothesis. Similarly, but in a statistical rather than
deterministic way, the bias can be defined in terms of {\it rms}
values of the galaxy and density fields
\begin{equation}
\left(\frac{\delta n}{\left<n\right>}\right)_{rms} = b \left(\frac{\delta\rho}{\left<\rho\right>}\right)_{rms} \,\,\,\,   .
\label{bias_rms}
\end{equation}
With this definition, $b$ is simply a ratio if the square root of
the {\it variance} of the galaxy and mass density fields measured over
a given scale $R$, i.e. $\sigma_R$(galaxies) = $b\; \sigma_R$ (mass). 
In our case, 
we can measure directly the relative bias between two populations $A$ and $B$, as

\begin{equation}
\frac{b_A}{b_B} = \frac{\sigma_8^A(r)}{\sigma_8^B(r)} \,\,\,\,  ,
\label{bias4}
\end{equation}
where $\sigma_8$ can be computed from the measured correlation
function as 
\begin{equation}
\sigma_R^2 = \frac{3}{R^3} \int_0^R r^2 \xi(r) dr\,\,\,\,  ,
\label{bias4a}
\end{equation}
which, for power-law form of $\xi(r)$, can be written as \citep{peebles80}
\begin{equation}
\sigma_8=\sqrt{C_\gamma\left(\frac{r_0}{8\ \rm{Mpc/h}}\right)^\gamma}\,\,\,\,  ,
\label{bias5}
\end{equation}
with
\begin{equation}
C_\gamma = \frac{72}{(3-\gamma)(4-\gamma)(6-\gamma)2^\gamma}\,\,\,\, .
\label{bias6}
\end{equation}

\begin{figure}	
  \includegraphics[width=9cm]{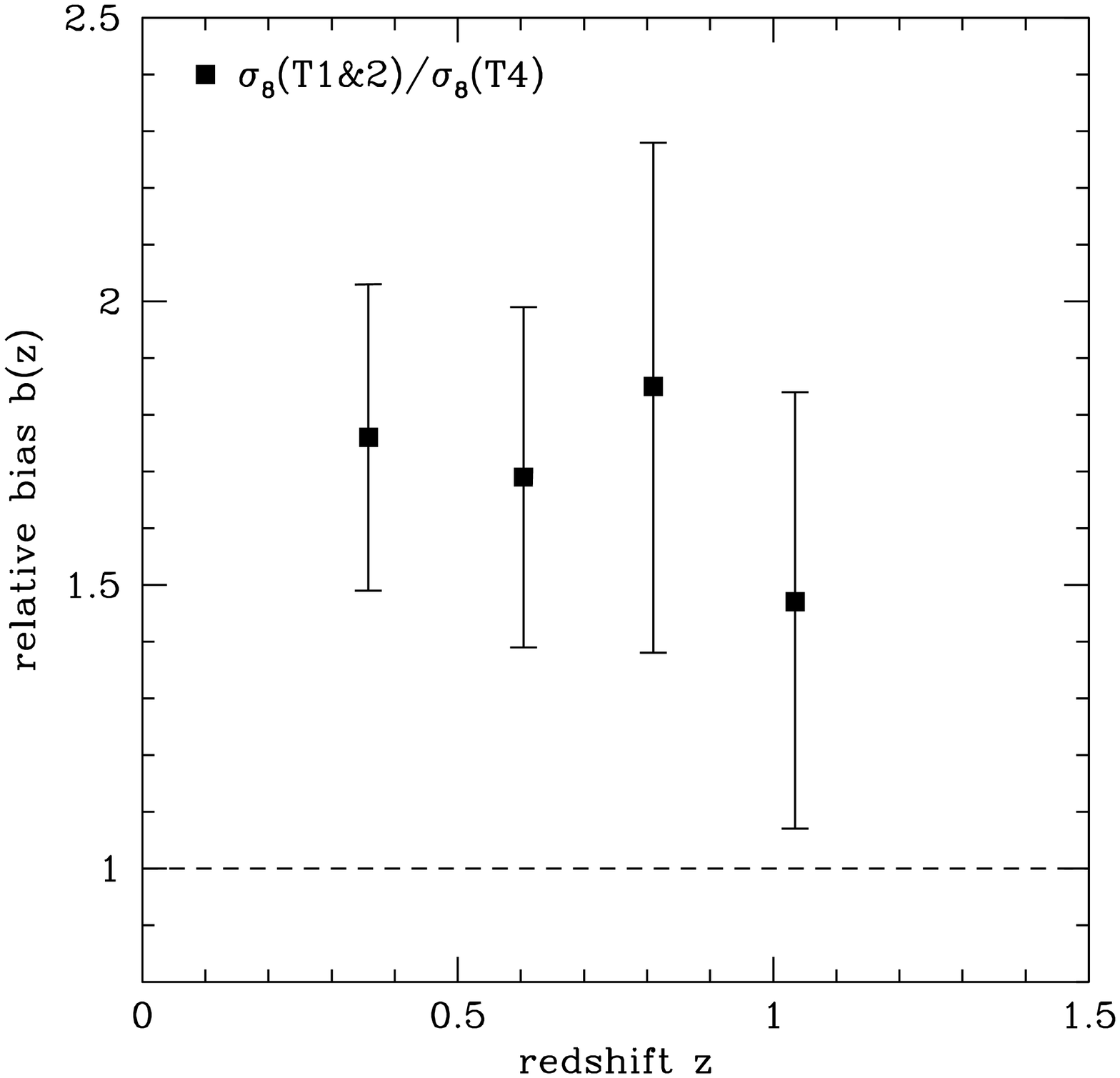}	
      \caption{Relative bias between populations of galaxies for a $8$~h$^{-1}$Mpc scale: $b_A/b_B=\sigma_8^A/\sigma_8^B$.}	
      \label{evol_bias}	
\end{figure}

Figure~\ref{evol_bias} shows the relative 
bias between the populations T1+T2 and T4 at different redshifts,
measured using eq.~\ref{bias5} in spheres with radius
$r=8$~h$^{-1}$Mpc.  The average value is 
$b=1.65 \pm 0.3$ consistent with being constant up to the highest redshift investigated.
This result seems to be mostly insensitive to the way the two 
populations are selected. For example we find a relative bias $b=1.6\pm0.3$ when
the red and blue samples are selected according to color bimodality (see next section).
Note, however that by considering the intermediate-type population in our comparison,
we find a relative bias $b=1.3\pm0.2$ between T1+T2 and T3, which is marginally
($\sim 1\sigma$) lower than between T1+T2 and T4.

This is in agreement with what is found locally for nearly the same color-selected
populations. For example \citet{wild2005} using the 2dFGRS find on the same
8~h$^{-1}$Mpc scale, $b \sim 1.5$.
Moreover, at higher redshifts, this result is in agreement
with what has been found in a previous 
VVDS study \citep{marinoni},
from the same data but using a different method.

\begin{table*}
\caption{Relative bias between early-type and late-type galaxies (from
spectral type classification), and 
``red'' vs. ``blue'' galaxies (separated using the color bi-modality)}
\label{table3_bias}
\centering
\begin{tabular}{l c c}
\hline\hline
                 & redshift  &         relative       \\
                 & range     &           bias         \\
\hline
T1\&2/T3         & [0.2-0.5] & $1.18^{+0.23}_{-0.22}$ \\ 
                 & [0.5-0.7] & $1.27^{+0.26}_{-0.28}$ \\ 
                 & [0.7-0.9] & $1.39^{+0.26}_{-0.27}$ \\ 
                 & [0.9-1.2] & $1.45^{+0.34}_{-0.37}$ \\ 
\hline
T1\&2/T4         & [0.2-0.5] & $1.76^{+0.27}_{-0.27}$ \\ 
                 & [0.5-0.7] & $1.69^{+0.30}_{-0.30}$ \\ 
                 & [0.7-0.9] & $1.85^{+0.43}_{-0.47}$ \\ 
                 & [0.9-1.2] & $1.47^{+0.37}_{-0.40}$ \\ 
\hline
``red''/``blue'' & [0.2-0.5] & $1.69^{+0.43}_{-0.47}$ \\
                 & [0.5-0.7] & $1.63^{+0.32}_{-0.34}$ \\
                 & [0.7-0.9] & $1.45^{+0.29}_{-0.30}$ \\
                 & [0.9-1.2] & $1.45^{+0.27}_{-0.27}$ \\
\hline
\end{tabular}
\end{table*}

\section{Rest-frame color bimodality} 

It has become clear in recent years that the galaxy rest-frame colors
show a bimodal distribution \citep{strateva}, i.e. that
in a (rest-frame) color-magnitude diagram
galaxies tend to segregate between a ``red sequence''
(similar, but less tight than that observed for cluster galaxies)
and a ``blue cloud''.   This behavior seems also to be
present at high redshift
\citep{bell2004,giallongo2005}.  From the SDSS data, \citet{strateva}
find that the blue cloud contains mainly late (spiral) morphological
types while the bulk of the red sequence consists of
bulge-dominated, early-type galaxies \citep[see also][]{weiner2005}.
This provides a natural recipe to split galaxies into two populations of
``red'' and ``blue'' objects and there have been already attempts to study the
clustering of galaxies separately for the two classes defined in this
way at intermediate redshifts \citep[\eg][]{phleps05}.

It is therefore interesting to check how
our analysis as a function of SED-defined types compares
to a simpler sub-division following the bimodal distribution.

\begin{figure}
  \includegraphics[width=9cm]{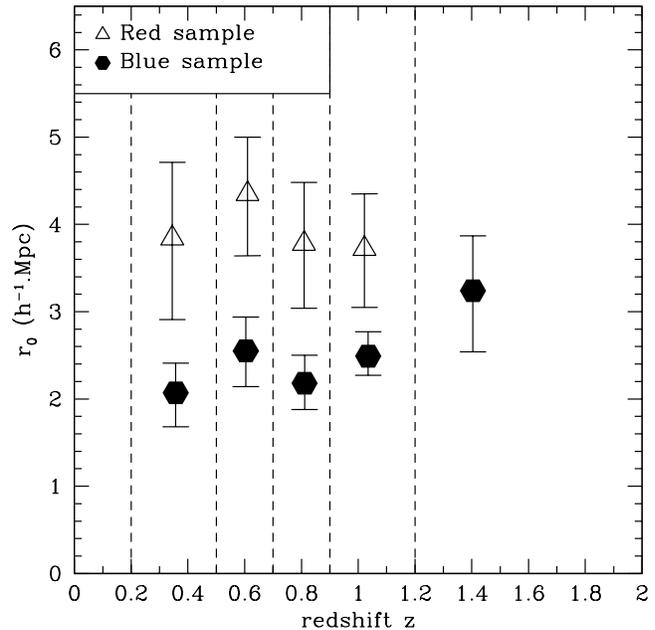}
  \caption{Correlation length $r_0$ as a function of redshift for the red and blue samples. 
    Here the slope $\gamma$ is a free parameter.}
  \label{evol_r0z_red_blue}
\end{figure}

We have thus split our data into a blue and red sample, following the rest-frame
color-magnitude relations suggested 
by \citet{giallongo2005} and measured the
projected function $w_p(r_p)$ and the best-fit power-law correlation
function $\xi(r)$ in different redshift bins.  
The resulting evolution of the correlation length 
is shown in Fig.~\ref{evol_r0z_red_blue}, with the measured
values for $r_0$ and $\gamma$  
summarized in Table \ref{table_red_blue}.
The associated errors have been computed in the usual way, constructing
blue and red mock catalogues from the
GalICS simulations, keeping the same proportion of red (blue) 
objects as in the real VVDS data.

Similarly to what we found for early-type classes, red-sequence
galaxies exhibit a larger clustering length ($r_0 \sim 4$~h$^{-1}$Mpc),
with little dependence on redshift.  Comparison 
to Fig.~\ref{evol_r0z} indicates that the classification into
rest-frame red and blue galaxies is substantially equivalent to
our classification into spectral late-types, producing samples
with similar clustering evolution properties.

The relative bias between red and blue galaxies is measured to be
on average $b=1.6\pm0.3$ between $z\sim0.3$ and $z\sim1$ (see Table \ref{table3_bias}),
a value comparable with that derived using the spectral type classification.

In summary, using the bi-modal split of the population we 
find results consistent with our analysis per spectral types. However,
we note that the simple rest-frame color-magnitude cut 
is a less precise way of selecting truly early-type galaxies, at least
for $z>0.7$, since
the full multi-color information is used to provide the spectral type
classification.

\begin{table*}
\caption{Number of galaxies and mean B rest-frame absolute magnitude (h=1)
of the sample separated using the color bimodality. 
The number of ``red'' and ``blue'' galaxies classified
as early (T1\&2) and late (T3\&4) spectral types are identified.
We note that the populations of ``red'' and T1\&2 as well
as the populations of ``blue'' and T3\&4 galaxies are
mostly the same, except for a significant cross-over population
which could explain the observed difference
in clustering (see text).}
\label{table_red_blue_t12_t34}
\centering
\begin{tabular}{c c c c c }
\hline\hline
 redshift  & \multicolumn{2}{c}{red} & \multicolumn{2}{c}{blue} \\
 range     &    T1\&2   &    T3\&4   &    T1\&2   &    T3\&4    \\
\hline
 0.2 - 0.5 &       221  &       49   &       86   &     895     \\
           &    -18.181 &   -16.403  &   -18.701  &  -17.411    \\
\hline
 0.5 - 0.7 &       253  &       24   &      121   &     1044    \\
           &    -18.892 &   -17.646  &   -19.215  &  -18.467    \\
\hline
 0.7 - 0.9 &       313  &       42   &       77   &     965     \\
           &    -19.713 &   -18.468  &   -19.748  &  -19.138    \\
\hline
 0.9 - 1.2 &       283  &       13   &       88   &     1016    \\
           &    -20.263 &   -19.314  &   -20.557  &  -19.679    \\
\hline
 1.2 - 2.0 &       41   &        0   &       94   &    480      \\
           &    -20.926 &        0   &   -21.078  & -20.567     \\
\hline
\end{tabular}
\end{table*}

\begin{table*}
\caption{Measurements of the clustering length and slope for 
galaxy subsamples divided following the color bimodality}
\label{table_red_blue}
\centering
\begin{tabular}{l c c c c c c}
\hline\hline
sample  & number of & redshift  & effective & $M_{B_{AB}} - 5 log(h)$ &         $r_0(z)$        &        $\gamma$        \\
        & galaxies  & range     & redshift  &        mean             &      (h$^{-1}$Mpc)     &                        \\
\hline
  red   &   270     & 0.2 - 0.5 &   0.345   &          -17.858        &  $3.84^{+0.87}_{-0.93}$ & $1.88^{+0.23}_{-0.16}$ \\
        &   277     & 0.5 - 0.7 &   0.610   &          -18.784        &  $4.35^{+0.65}_{-0.71}$ & $1.99^{+0.22}_{-0.18}$ \\
        &   355     & 0.7 - 0.9 &   0.809   &          -19.565        &  $3.78^{+0.70}_{-0.74}$ & $1.87^{+0.28}_{-0.22}$ \\
        &   296     & 0.9 - 1.2 &   1.022   &          -20.221        &  $3.72^{+0.63}_{-0.67}$ & $1.83^{+0.32}_{-0.32}$ \\
\hline
  blue  &   980     & 0.2 - 0.5 &   0.357   &          -17.521        &  $2.07^{+0.34}_{-0.39}$ & $1.67^{+0.14}_{-0.09}$ \\
        &  1164     & 0.5 - 0.7 &   0.604   &          -18.544        &  $2.55^{+0.39}_{-0.41}$ & $1.68^{+0.15}_{-0.09}$ \\
        &  1042     & 0.7 - 0.9 &   0.811   &          -19.183        &  $2.18^{+0.32}_{-0.30}$ & $1.40^{+0.16}_{-0.11}$ \\
        &  1105     & 0.9 - 1.2 &   1.035   &          -19.749        &  $2.49^{+0.28}_{-0.22}$ & $1.84^{+0.14}_{-0.10}$ \\
        &   574     & 1.2 - 2.0 &   1.404   &          -20.651        &  $3.24^{+0.63}_{-0.70}$ & $1.91^{+0.20}_{-0.18}$ \\
\hline
\end{tabular}
\end{table*}


\section{Discussion and conclusions}

The fundamental result of this investigation is to have
clearly established that at least up to $z\simeq1.2$ 
early-type galaxies continue to be more strongly clustered than late-type
galaxies. This extends to higher redshifts earlier results obtained
in the local Universe \citep{loveday,norberg,zehavi2002,zehavi2005}, 
at redshifts $z\simeq0.6$ \citep{phleps05}, and quantitatively
confirms early indications at $z\simeq1$ \citep{phleps}. 

Our specific results can be summarized as follows:\\
1. The clustering of sub-L$^*$ early-type galaxies is stronger than the
clustering of late-type galaxies up to $z\simeq1.2$; a similar
behavior is obtained for ``red-sequence'' and ``blue-cloud''
galaxies selected using rest-frame colors.\\
2. Luminous, early-type galaxies are already strongly clustered at
$z\sim1$ with {\bf 3.5}$\leq r_0 \leq 5$~h$^{-1}$Mpc.\\
3. The clustering of late-type galaxies is low at all epochs
with $r_0\simeq2.1-3.5$~h$^{-1}$Mpc.\\
4. The overall clustering of bright late-type galaxies at $z\simeq1.4$
is relatively strong with $r_0=3.4$~h$^{-1}$Mpc \\
5. The bias between the early-type and late type
population is nearly constant, $b_{early}/b_{late}=1.7$ over the redshift
range $0.2 \leq z \leq 1.2$.

A physical interpretation of these results and comparison to
local values require to keep
in mind that we are working with a flux-limited sample,
and thus that we are measuring galaxies of different intrinsic
luminosity at different redshifts.  At $z\sim1$ the VVDS targets early-type galaxies 
with $-22 \leq M_B(AB) \leq -19$ and an average $M_B(AB)=-20.3$
(see Figure \ref{brest}). Early-type galaxies observed in the 2dFGRS 
with similar luminosities have an observed $r_0=6-7.5$~h$^{-1}$Mpc,
increasing with luminosity \citep{norberg}.   
The clustering amplitude of $\xi(r)$ 
for early-type galaxies has therefore evolved by no more than 
70\% between redshift $\sim1$ and present.  

\begin{figure*}
\begin{center}
  \includegraphics[width=13cm]{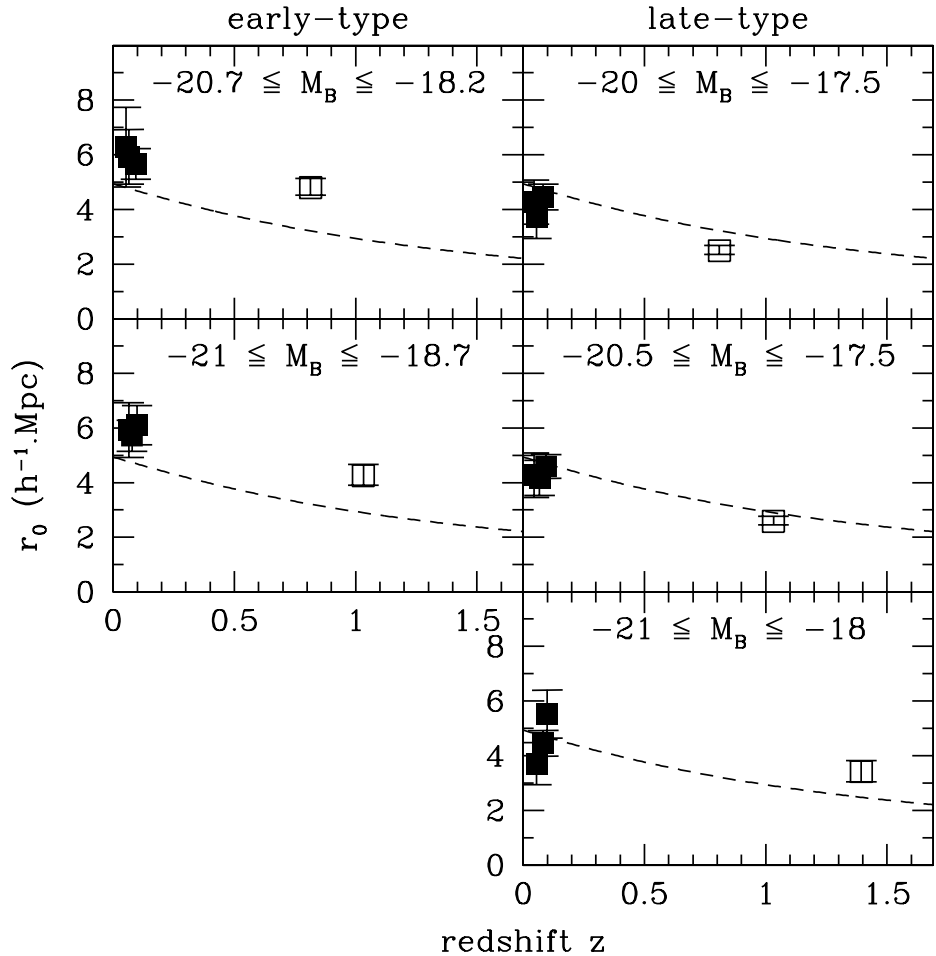}
  \caption{Clustering length evolution for early (left panel)
    and late (right panel) spectral type galaxies.
    Measurements from the VVDS at high redshifts (open symbols) are
    compared to the local measurements obtained for galaxies in the
    same absolute luminosity ranges in the 2dFGRS (filled symbols) surveys
    \citep{norberg}, with an evolution correction of 0.5 magnitudes for early types
    and 1 magnitude for late types at $z\sim1$ \citep{zucca}. 
    The dashed curves show the predicted gravitational growth of the dark matter,
    with amplitude normalized to $\sigma_8=0.9$ at the current epoch.}
  \label{corev}
\end{center}
\end{figure*}

To understand the
implications of this observation, we have compared our observed evolution of $r_0$
to the predicted evolution of the clustering length for the dark matter in a $\Lambda$
cosmology (using the expected evolution of the variance as described in
\citet{peebles80}). As shown in Figure~\ref{corev},  if one assumes that the
early-type population is simply tracing in an unbiased way the growth of clustering in the mass
(as would have been the case if no evolution in number nor in luminosity occurred) 
their clustering would increase
by a factor of
$\sigma_8(z=1)/\sigma_8(z=0)\simeq0.6$, which is close to what 
we see in Figure \ref{corev}.
This result is in agreement with the observation of a marginal
clustering evolution of early-type galaxies since $z\sim1.5$
\citep{cimatti2004,saracco2004}, i.e. with the idea
that the bulk of the spheroid population was already in place at
these early redshifts. If a significant fraction of these
galaxies was assembled and added below $z\sim 1$, we would observe
not only an increase in their number density, but also a weakening of
their clustering signal with respect to the underlying mass: it is
natural to think that, forming
typically in lower density peaks than their high-redshift companions
they would in fact be less massive and less clustered.

The lower clustering level observed for early-type galaxies at
$z<0.7$, is best interpreted as a luminosity effect, 
given the faint typical luminosity of galaxies in the first bins
(\eg $\left<M_B(AB)\right>=-18.3$ at  $z=[0.2,0.5]$).

The trend observed for blue star-forming galaxies of similar luminosities
is also very interesting.
Below redshift $z\simeq1$, their clustering 
is observed to be weak and continuously decreasing from high
redshift to the present.  As shown in the two right-top panels of
Figure~\ref{corev}, however, this is consistent with simple
gravitational growth: the clustering amplitude measured at $z\sim 1$
for these galaxies is coherent with the evolution of the same
structure traced by the same galaxies at $z\sim0$.
The point at $z=1.4$ may indicate that star formation activity at $z>1$ 
is found in progressively more strongly clustered galaxies than at present,
i.e. plausibly in more massive galaxies in higher density peaks.
In our sample, galaxies at $z=[1.2,2]$ have 
a mean $M_B(AB)=-20.5$ and a clustering length
$r_0=3.5$~h$^{-1}$Mpc. Taking into account 1.5 magnitudes of
B-band brightening \citep{zucca} these galaxies would have today
$M_B(AB)=-19$. Late-type galaxies with similar luminosities in the
Stromlo-APM, 2dFGRS or SDSS have $r_0\simeq2.9-4.2$~h$^{-1}$Mpc
\citep{loveday,norberg,zehavi2002}. The clustering amplitude of 
late-type star-forming galaxies with comparable luminosities, therefore 
remains roughly constant since $z\sim 1.5$ (Figure \ref{corev}).
A straightforward explanation to produce the observed result is that as
redshifts get smaller, star formation is shifting to 
smaller-mass, less clustered galaxies. 

It is difficult to compare correlation properties of samples with different
mean/median magnitudes because of the well known dependence of clustering
on luminosity \citep{guzzo1997,budavari2003,zehavi2002,pollo2006}. Nevertheless, we can note that
comparing our values with other
clustering measurements for type- or color- selected galaxies at
redshifts beyond 1.5, we find for example that star-forming LBGs
at $z\sim 3$ \citep{adel} are even more clustered
($r_0\sim 4.5$~h$^{-1}$Mpc) than our star-forming galaxies at $z\sim1.5$.
This is consistent with the VVDS results at lower redshifts.
One can speculate that star formation is moving to higher and
higher density environments, when moving back in time.
Similarly, red selected samples of EROs
at $z\sim1.5$ or red K-selected
galaxies with $2 < z_{phot} < 4$
are significantly more clustered than the early type
galaxies at $z\simeq1$ in the VVDS with
$r_0$ up to $8$~h$^{-1}$Mpc \citep{daddi2002,daddi}. These
galaxies seem to be populating even more 
extreme density peaks than those we have
been studying here at $z\sim 1$.


\begin{acknowledgements}
We thank the referee, Richard Kron, for very useful comments.\\
This research program has been developed within the 
framework of the VVDS consortium.\\
This work has been partially supported by the
CNRS-INSU and its Programme National de Cosmologie (France),
and by Italian Ministry (MIUR) grants
COFIN2000 (MM02037133) and COFIN2003 (num.2003020150).\\
The VLT-VIMOS observations have been carried out on guaranteed
time (GTO) allocated by the European Southern Observatory (ESO)
to the VIRMOS consortium, under a contractual agreement between the
Centre National de la Recherche Scientifique of France, heading
a consortium of French and Italian institutes, and ESO,
to design, manufacture and test the VIMOS instrument.
\end{acknowledgements}


\end{document}